\documentclass[12pt]{article}
\usepackage[utf8]{inputenc}
\usepackage[english]{babel}
\usepackage{amsthm}
\usepackage{graphicx}

\usepackage{algorithm}
\usepackage{algpseudocode}
\usepackage{dsfont}
\usepackage{amssymb}
\usepackage{float}
\usepackage{xcolor}
\usepackage{amsmath}

\DeclareMathOperator*{\argmax}{arg\,max}
\usepackage{graphicx}
\usepackage{subfig}

\usepackage[round]{natbib}
\usepackage{amsmath}   

\begin{document}

\begin{center}
{\Large\bf A method to identify geochemical mineralization on linear transect}
\end{center}
\begin{center}
{\sc Dominika Mikšová, Christopher Rieser, Peter Filzmoser}\\
{\it TU Wien, Vienna, Austria}\\
e-mail: {\tt dominika.miksova@tuwien.ac.at}\\

{\sc  Simon M. Thaarup}\\
{\it GEUS, Denmark\\}

{\sc  Jeremie Melleton}\\
{\it BRGM, France\\}
\end{center}

\begin{abstract}
    Mineral exploration in biogeochemistry is related to the detection of
anomalies in soil, which is driven by many factors and thus a complex
problem.
\citet{miksova} have introduced a method for the identification of spatial 
patterns with increased element
concentrations in samples along a linear sampling
transect. 
This procedure is based on fitting Generalized Additive Models (GAMs) 
to the concentration data, and computing a curvature measure from the 
pairwise log-ratios of these fits. The higher the curvature, the more likely 
one or both elements of the pair indicate local mineralization.
This method is applied on two geochemical data sets which have been 
collected specifically for the purpose of mineral exploration. 
The aim is to test the technique for its ability to identify 
pathfinder elements to detect mineralized
zones, and to verify whether the method can indicate which sampling material 
is best suited for this purpose.
\end{abstract}








\section{Introduction}

The identification of mineralized zones belongs to the important challenges in applied geochemistry.
The difficulty is that the targeted mineralizations could be of any arbitrary size, and in any
depth, depending on the type of mineralization. The common procedure to discover mineralized zones
is based on sampling, using strategic sampling designs in order to be as economic as possible.
Samples can be taken from different soil layers, but also from different trees and plants 
around the presumed target. Since samples and their analysis of element concentrations is
cost intensive, the sampling is often done on linear transects, crossing the presumed 
mineralized zones. If there is more evidence, drilling is also used in order to obtain a 
depth profile of the element concentrations. More information on different sampling strategies
can be found in \citet{report}.

In this work we assume that the sampling has been carried out on a linear transect, or that 
the available samples can be aggregated to such linear transects. This means that the spatial
locations can be considered along a line, and thus it is simple to graphically investigate the 
spatial variability of the measured elements by simply plotting the element concentrations
against the locations \citep{lineplot}.
However, in modern geochemistry, the number of elements that can be reliably measured is in
the range of 30-60, and if the samples have been obtained from several different sampling media,
it is a challenging task to study all resulting plots for abrupt changes in the concentration
values. Such changes could indicate mineralized zones, since their signals could lead to
sudden increases of element concentrations. There is, however, the problem
that due to the (economic) sampling procedure, only very few samples might have been taken
on top of the mineralizations, and together with measurement and analysis uncertainties, the 
resulting concentration changes might not be clearly expressed. The second problem is that 
there is an interplay of the concentration values among the elements, because geochemical data
are compositional by their nature \citep{Aitchison86,FHT2018}.

Consider a composition $x_1^m, \ldots ,x_{D_m}^m$, consisting of $D_m$ chemical elements,
measured in $m=1,\ldots ,M$ different sample materials. For the analysis of compositional data 
it has become popular to use the so-called log-ratio methodology, introduced by
\citet{Aitchison86}. This refers to the use of logarithms of ratios, and the basic information
are log-ratio pairs $\ln (x_j^m / x_l^m)$, for $j,l \in \{1,\ldots ,D_m\}$. The use of log-ratios
eventually leads to a sound geometrical concept, referred to as the Aitchison geometry \citep{pawlowsky15}.
There are, however, also practical reasons why log-ratios are useful, such as symmetry around zero,
and equal variance if numerator and denominator are exchanged.

A further argument for considering (log-)ratios is the assumption that there could be elements
which are stable and thus not affected by a mineralization, and others are very indicative of
mineralized zones. The log-ratio of such elements could even better express the local change around
a mineralization, because measurement and analysis uncertainties could cancel each other out \citep{miksova}.

On the other hand, $D_m$ different elements would lead to $D_m(D_m-1)/2$ different (and relevant)
pairwise log-ratios, which makes a visual inspection practically impossible. For this reason,
\citet{miksova} have introduced a procedure to rank the list of log-ratio pairs according to 
their ability to indicate mineralization. This is done by first approximating the individual 
element concentration by a smooth fit, taking log-ratios of the smooth fits, and computing 
a measure of curvature. The higher the curvature, the more likely (at least) one of the log-ratio pair 
elements shows sudden changes. In addition, the visualization of the smooth fits and their log-ratio
allows to localize the presumed mineralized zones.

In this paper we briefly review the method of \citet{miksova}. Then we apply this procedure to 
two geochemical data sets, originating from surveys carried out in Greenland and France, respectively,
in the frame of the ongoing project ``UpDeep'' \citep{updeep}, which aims
at developing
and implementing a methodology to identify mineralization.

\section{Methodology}

As already indicated in the introduction, the main idea of the methodology developed in \citet{miksova} is that at the beginning and end of a transect crossing
a potential mineralization, important log-ratios of an element pair display a very quick spatial change which can be captured by a measure based on the curvature of the latter. 

The first step of the methodology consists in fitting a so called GAM model, see \citet{gamWood}, to each element, with concentration values $y_i$ at 
locations $x_i$, for $i=1,\ldots ,n$. After considering the nature of our data and after  inspection of the corresponding residual plots we decided to model the data belonging to the Tweedie family with a log-link and additional weights. Modelling the element concentrations in such a way means that for each element the following optimization problem based on the log-likelihood function $l$ is solved to obtain a linear predictor $\eta$
\begin{align*}
    \hat{\eta} = \argmax_{\eta \in \mathcal{H}} \, \sum^{n}_{i=1} \omega_i l(y_i|x_i;\eta) - \lambda \int (\eta''(x))^2 dx,
\end{align*} 
with predefined weights $\omega_i$, upweighting certain points, a suitable function space $\mathcal{H}$ and a smoothing parameter $\lambda$. 

This results in GAM fits $\hat{f}_{el_1}$ and $\hat{f}_{el_2}$
for each pair of elements $el_1$ and $el_2$, and 
the log-ratio of the fits for any location $x$ along the transect
can be obtained subsequently as 
\begin{align*}
 g(x):&=\log\bigg(\frac{\hat{f}_{el_1}(x)}{\hat{f}_{el_2}(x)}\bigg)\\ &= \log\big(\hat{f}_{el_1}(x)\big) - \log\big(\hat{f}_{el_2}(x)\big) \\
 &= \log\big( h^{-1}(\hat{\eta}_{el_1}(x)) \big) - \log\big(h^{-1}(\hat{\eta}_{el_2}(x))\big),
\end{align*}
where $h(\cdot)$ stands for link function. 

Its curvature is then computed by  
\begin{align*}
        \kappa(x):= \frac{|kg''(x)|}{(1+(kg'(x))^2)^{\frac{3}{2}}},
\end{align*}
where $k$ is a scaling factor allowing the curvatures to be comparable amongst different pairs.

Finally, for each pair of log-ratios, the following quantity is introduced to measure quantitatively important spatial changes potentially indicating the
beginning and the end of a mineralization, namely:
\begin{equation}
\label{eq:cvalue}
    c(el_1,el_2):= \frac{2}{L} \sum^{\frac{L}{2}}_{l=1}  \max_{x \in [x_{j_{2l-1}},x_{j_{2l}}]} (\kappa(x)-\mathcal{T})_{+}^2. 
\end{equation}
This measure is denoted as the $c$-value in the following.
Here, $(\cdot)_+$ denotes $\max(\cdot,0)$, and $\mathcal{T}$ is a threshold, $L$ is the number of times that the curvature $\kappa$ crosses the threshold, and $[x_{j_{2l-1}},x_{j_{2l}}]$ are the corresponding points where this happens. It is easy to see that only points $x$ for which the curvature is above the threshold are influencing this measure $c(el_1,el_2)$. This avoids any influence of small values of $\kappa(x)$, meaning that only very high signal changes of the log-ratio are taken into account. Summing up over all maximum leads to a quantity measuring the mean number of high signal changes.

For a more detailed description of the weights $\omega_i$, the smoothing parameter $\lambda$, the scaling factor $k$, the threshold $\mathcal{T}$, and the numerical computation of the derivative, as well as the measure $c(el_1,el_2)$ we refer to \citet{miksova}.

Since we are dealing with compositional data, one could argue that not the absolute element concentrations
should be used for the GAM fits, but rather the log-ratios of all pairs of elements. Although this could be a
reasonable approach, there are several arguments against this idea: (a) The GAM fits may require some manual adjustment
and tuning, which is not feasible for all pairwise log-ratios. (b) Typically, the number of observations is rather
low, and there could be some data quality issues as well. GAM fits on the raw data could, to some extent, ``repair''
this effect, particularly if there is uncertainty in small concentrations, and ideally the data quality after the
log-ratios of the GAM fits increases.

\section{Results}

One important part of the UpDeep project has been to take samples in two countries, namely in Greenland and France. This sampling procedure was successfully accomplished by the sample providers GEUS (Geological Survey of Denmark and Greenland) and BRGM (The French Geological Survey), respectively. The sampling was performed by executing geochemical sampling surveys according to the established protocols in geologically well-known mineralized areas. The samples in both countries were taken in the years 2017 and 2018, however in this context we focus on one specific year 2018. 


\subsection{GEUS data}

\begin{figure}[!ht]
\begin{center}
\includegraphics[width = 0.7\linewidth]{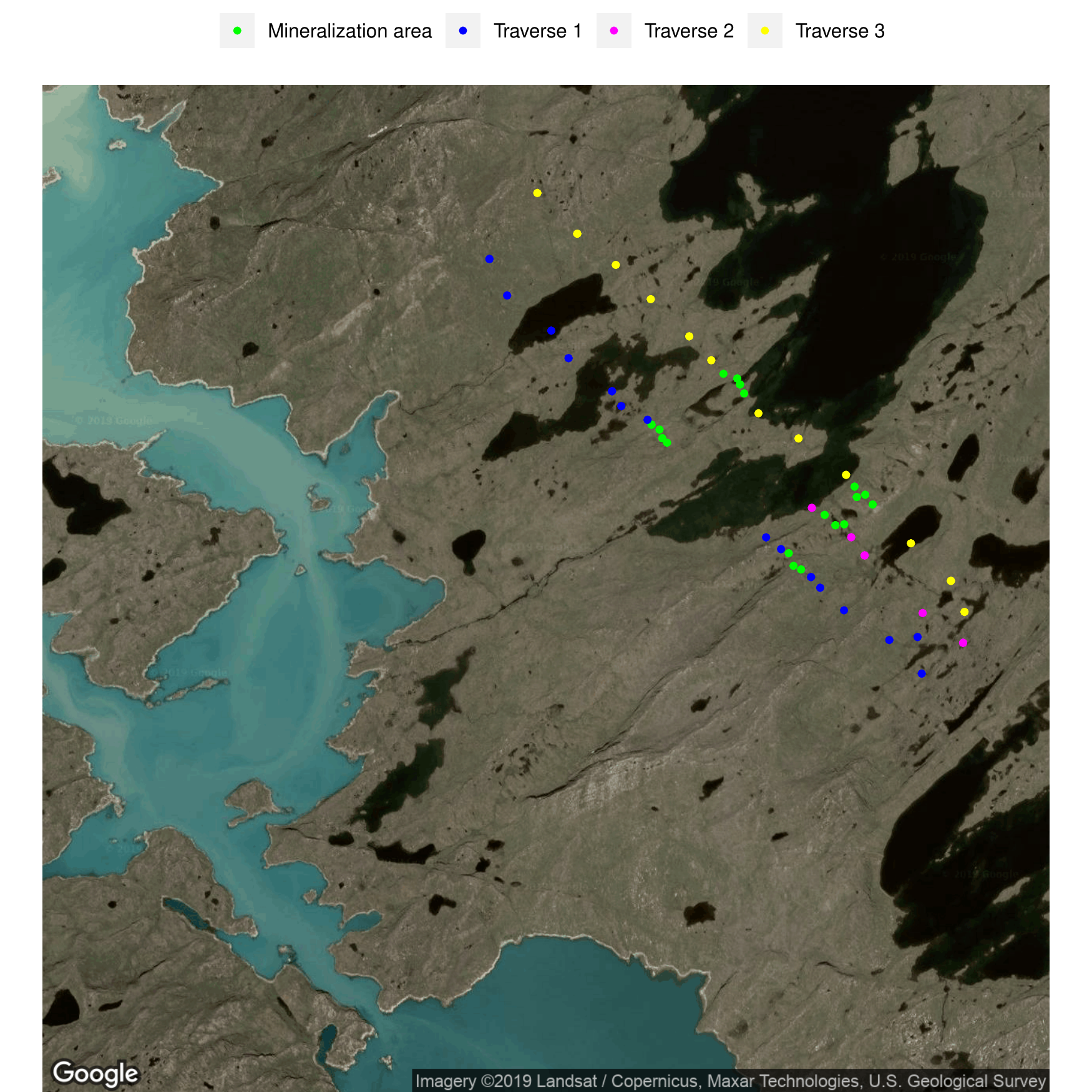}
\caption{\label{fig:GEUSmap}Map of the locations of the samples taken 
by GEUS in the Isortoq South Area.}
\end{center}
\end{figure}

The sampling areas were chosen due to known mineralization and exploration in the area. 
The interest is in the area Isortoq, which is situated in the very south of Greenland, see Figure~\ref{fig:GEUSmap} for a detailed map, where the
map background is obtained using Google maps \citep{googlemap}. 
In total, three traverses were sampled which are 300 meters apart.
The samples from the different traverses are shown in different color in the
map. Green color refers to the locations of known mineralizations.
In this case the deposit is an Iron (Fe), Vanadium (V), Titanium (Ti) deposit. A 
possible proxy for V could be Scandium (Sc) (since V tends to be analyzed poorly).
In our analyses we merge the samples from the three traverses into one linear
transect, which means that all
samples have been taken, but their locations are set to a linear transect in the 
center of the three traverses.
The individual samples are now at a distance between 50 to 400 meters. 
The total length of the transect is about 12 km.

Two different plant species and soil samples have been investigated, namely 
Salix Glauca and Empetrum Nigrum with 49 samples, and soil comprises 47 observations containing only so called routine samples.

\begin{figure}[!ht]
\begin{center}
\includegraphics[page={3}]{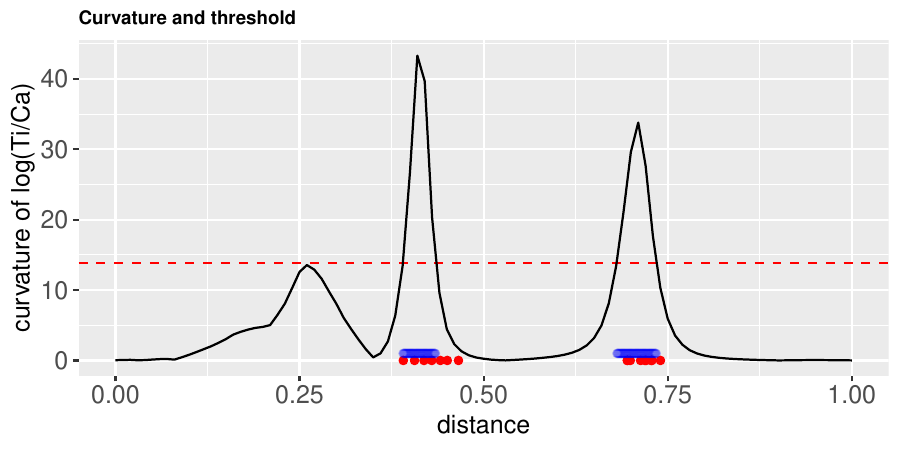}
\caption{\label{fig:geus_min_Ti}Curvature of the log-ratio of the GAM fits
of Ti (Titanium) and Calcium (Ca) in soil.}
\end{center}
\end{figure}

\begin{figure}[!ht]
\begin{center}
\includegraphics[page={15}]{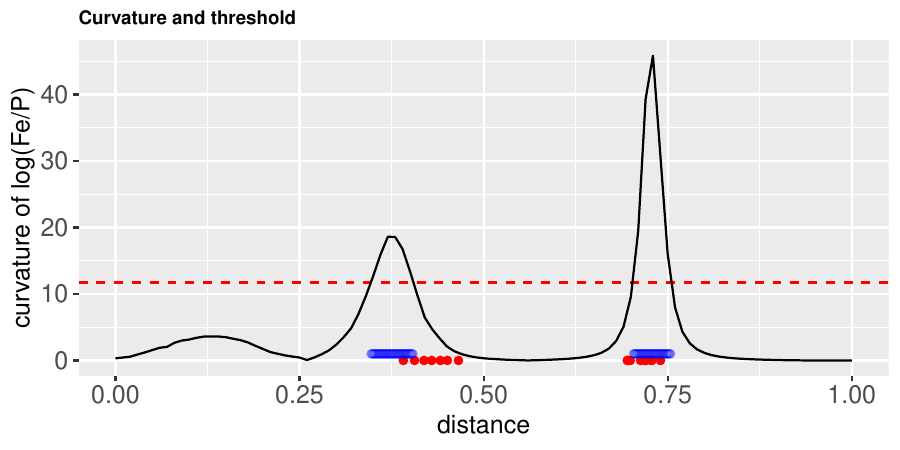}
\caption{\label{fig:geus_min_Fe}Curvature of the log-ratio of the GAM fits
of Fe (Iron) and P (Phosphorus) in soil.}
\end{center}
\end{figure}

Following the procedure of Section~2, Figure~\ref{fig:geus_min_Ti} shows 
the log-ratio pair of the GAM fits of the elements Ti and Ca (Calcium)
measured in soil, and Figure~\ref{fig:geus_min_Fe} displays the resulting
log-ratio for Fe and P (Phosphorus) in soil. Both log-ratios yield 
top-ranked $c$-values, see Equation~(\ref{eq:cvalue}).
In these plots, the mineralized zones are shown by red points. The blue
points are the predicted mineralizations, when the curvature exceeds the
threshold $\mathcal{T}$, which is indicated by the horizontal dashed line.
The predictions confirm the presumed mineralized zones very well, and they 
do not indicate new mineralized areas.

\begin{figure}[!ht]
    \centering
    \subfloat[Salix Glauca]{{\includegraphics[page = {2}, width=7.1cm]{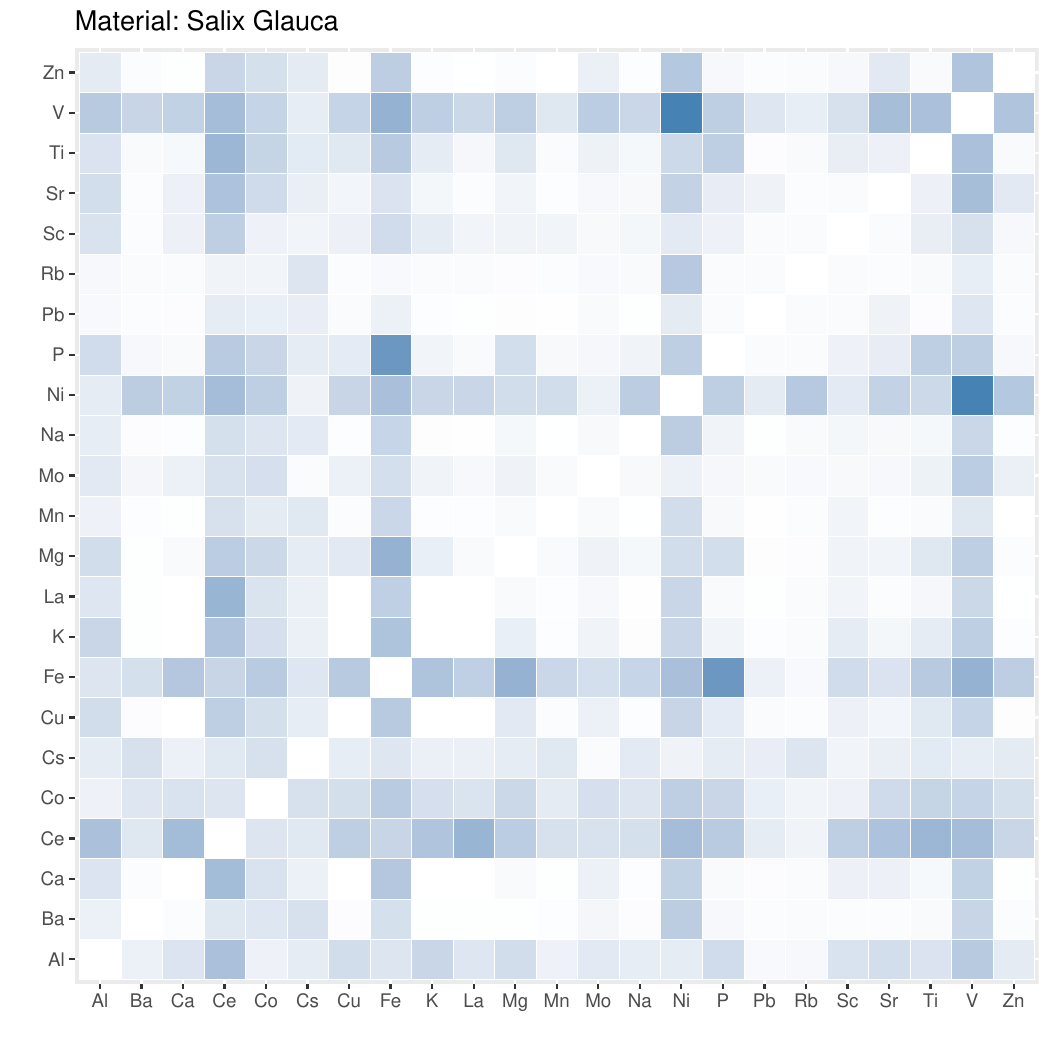} }}%
    \qquad
    \subfloat[Empetrum Nigrum]{{\includegraphics[page = {3}, width=7.1cm]{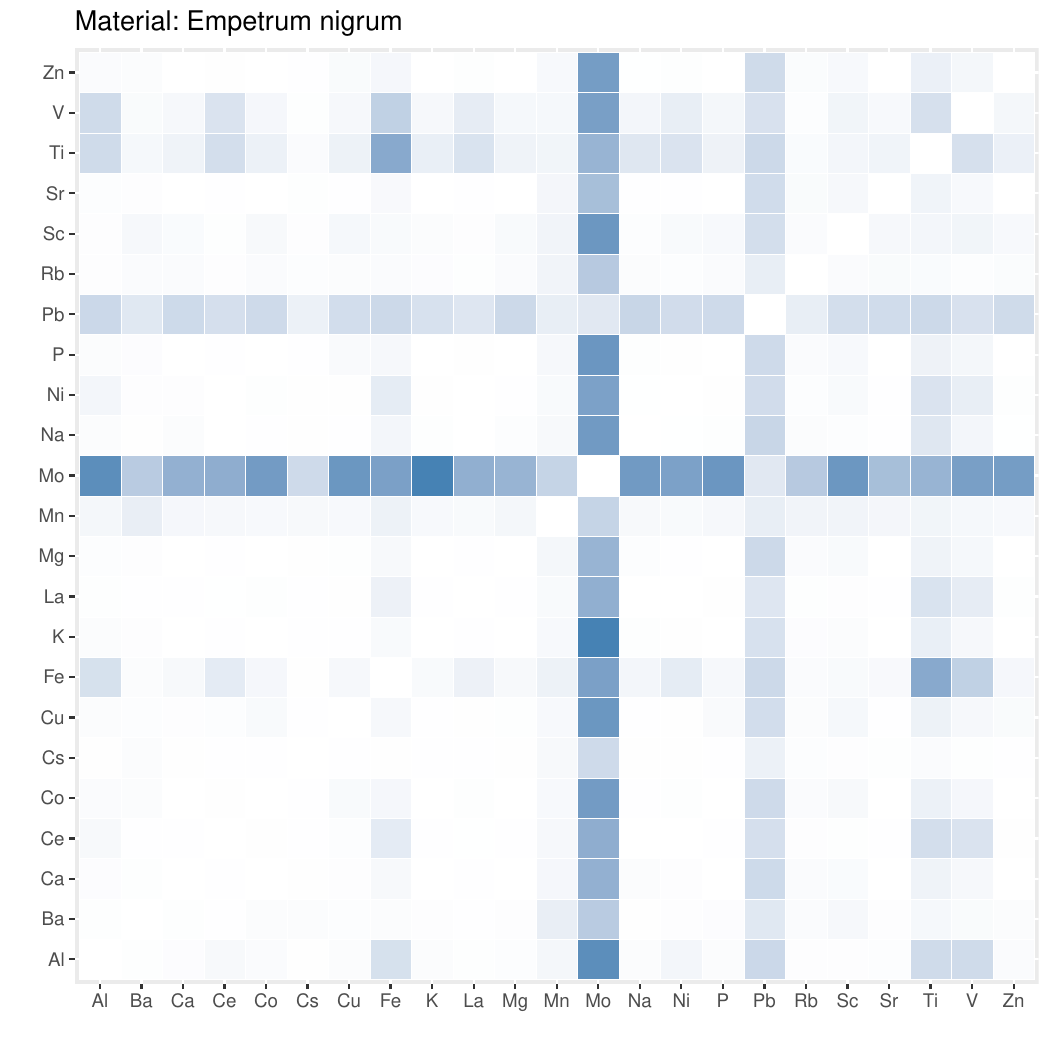} }}
    \qquad%
    \subfloat[Soil]{{\includegraphics[page = {2}, width=7.1cm]{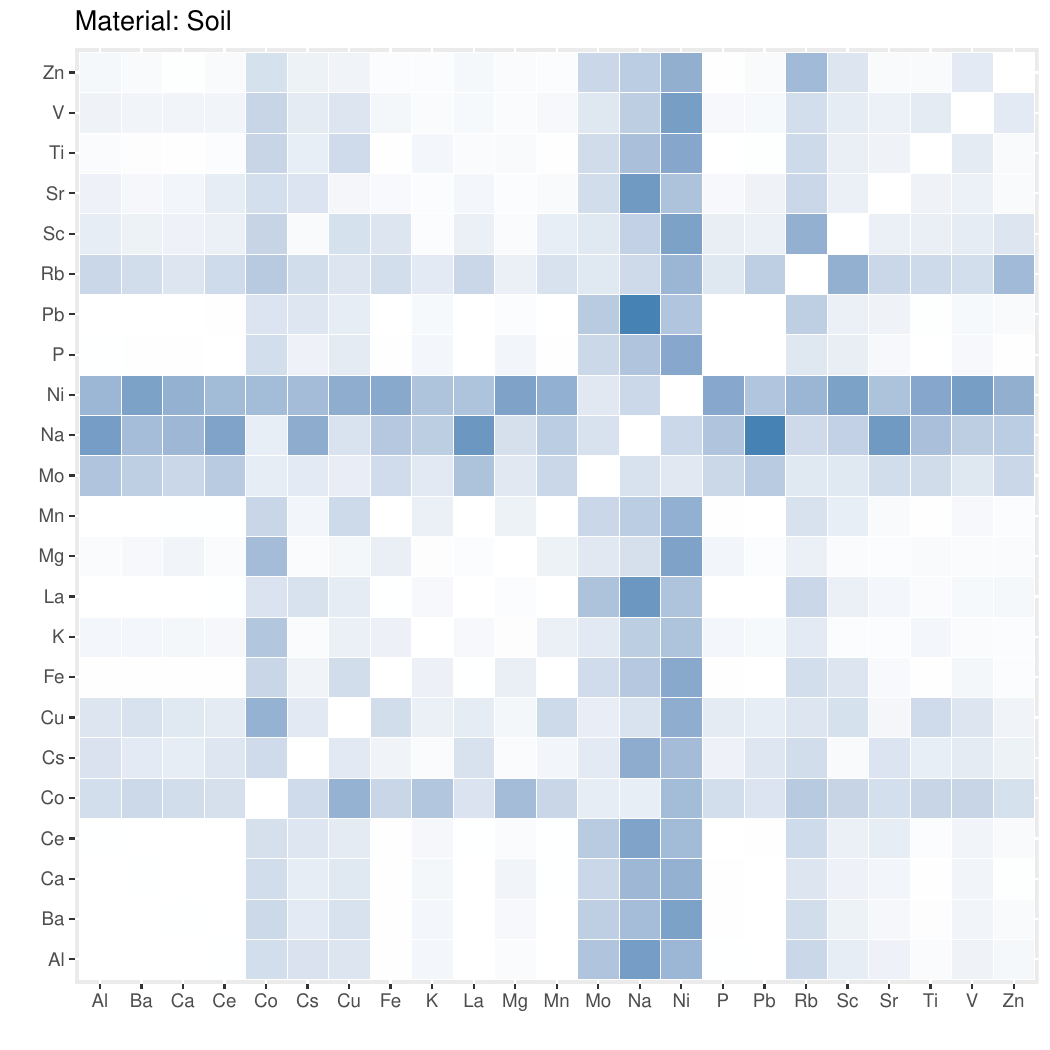} }}%
    \caption{Heatmaps of the $c$-values for the plant materials and soil.}
    \label{fig:heat_geus_plant}%
\end{figure}

A useful tool to display the overall information about meaningful log-ratios 
is the heatmap. The input for the heatmap is a matrix of $c$-values, computed 
from the GAM fits of all
pairwise log-ratios of a specific sampling material (plants or soil).
Figure~\ref{fig:heat_geus_plant} shows the resulting heatmap for
Salix Glauca (left), Empetrum Nigrum (right), and soil (bottom). 
Obviously, the heatmaps are symmetric due to the symmetry of the log-ratios.
The darker the blue color, the higher is the $c$-value obtained from the
corresponding log-ratio. A dark blue row or column in the heatmap indicates
so-called pathfinder elements, which potentially refer to mineralization.
From a geochemical point of view, most of the elements with higher
$c$-values are related to the deposit (Fe-V-Ti).

\begin{figure}[!ht]
\begin{center}
\includegraphics[page={16}]{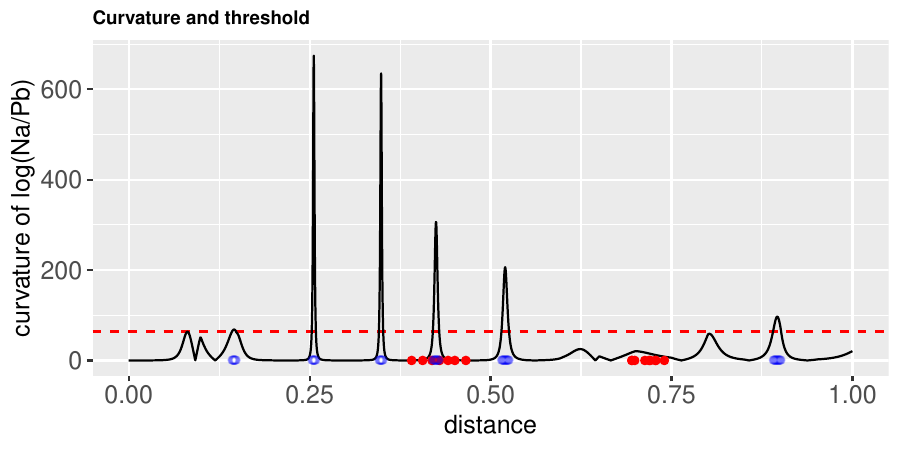}
\caption{\label{fig:geus_min_Na}Curvature of the log-ratio of the GAM fits
of Na (Sodium) and Pb (Lead) in soil.}
\end{center}
\end{figure}

These heatmaps can also be used to identify potentially interesting 
pathfinder elements that could indicate new mineralized zones. 
For example, the heatmap for soil 
(bottom plot in Figure~\ref{fig:heat_geus_plant}) shows a high
$c$-value for the pair Na (Sodium) and Pb (Lead).
Figure~\ref{fig:geus_min_Na} presents the corresponding curvature plot
of this log-ratio. Indeed, there are several locations where
abrupt signal changes are visible. One would have to further explore
these locations.


\subsection{BRGM data}

\begin{figure}[!ht]
\begin{center}
\includegraphics[width = 0.7\linewidth]{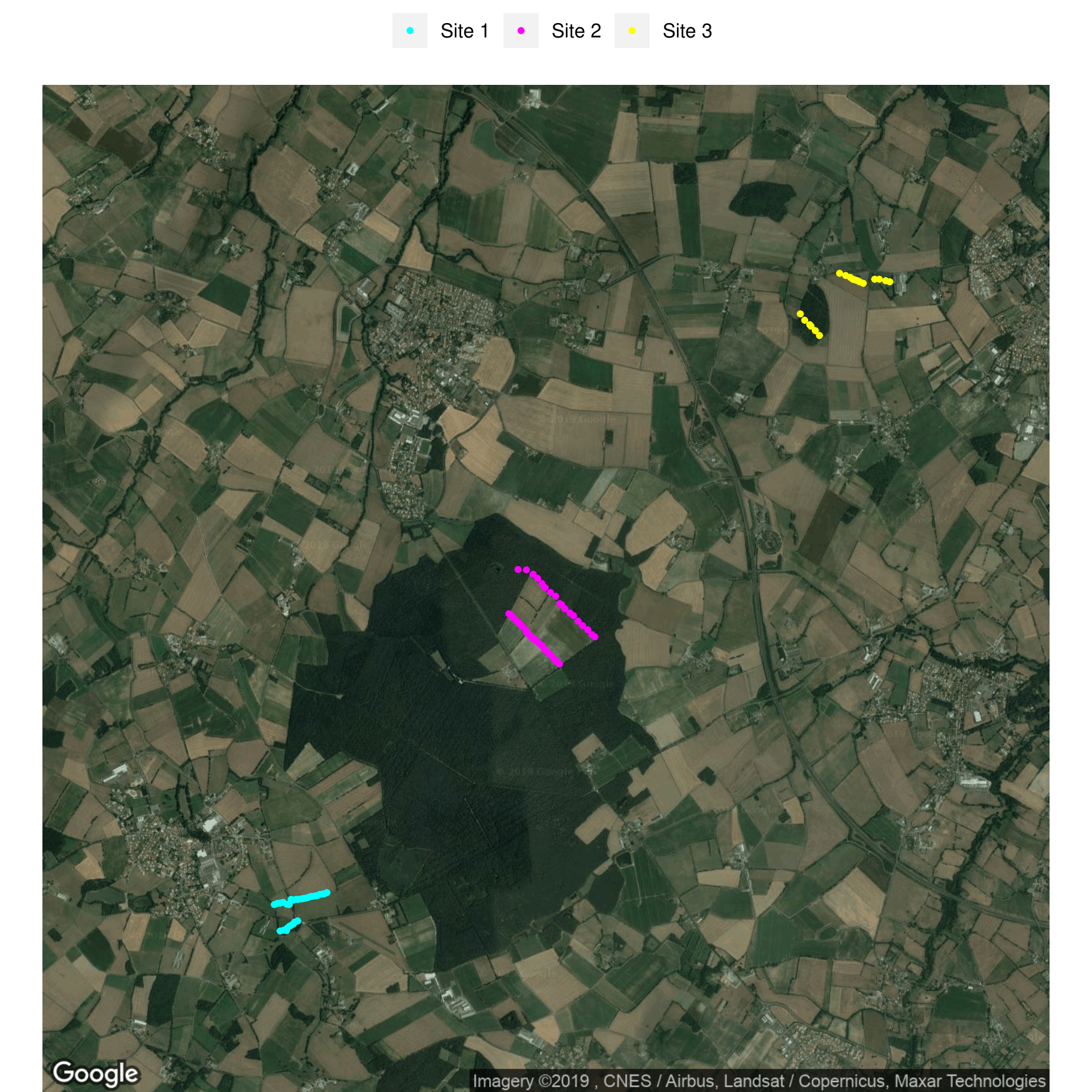}
\caption{\label{Francemap}Map with the sample locations taken by BRGM in the Vend\'ee area in 2018}
\end{center}
\end{figure}

The second data set originates from the Vend\'ee area in middle-west France,
which has been sampled in 2018. The area was investigated because of
some historical knowledge of the occurrence of rare elements.  
Moreover, an easy access allowed for a valuable recognition of the area 
prior to sampling. Figure~\ref{Francemap} shows a satellite map of 
the area where the samples have been taken from three different 
sites. Each of these subareas contains two traverses which are again
merged to one transect in our procedure in order to increase the 
number of observations per site.
The first site in the south-west of Figure~\ref{Francemap} holds 
approximately 30 samples, the second (middle) site about 40, and the 
third (north-eastern) site only 18 samples. The presumed mineralization type 
on all sites is Antimony (Sb) and Gold (Au). Due to pre-studies it turned 
out that the second site has the highest concentrations of Sb. The element
Au is in any case difficult to measure.

This data set provides in total 6 different sample materials, namely 
Ah horizon with Aqua Regia leach (AhAQ), Ah with deionized water leach (AhL1), 
Ah with sodium pyrophosphate leach (AhL3), Bramble branch (BB), Bramble leaves (BL), and Oak bark (OB). 
Rather than investigating again the curvature plots, we focus now on the
task to identify the most promising sample material indicating mineralization.
An answer would be highly relevant, because sampling of the different materials
is very time- and cost-intensive.

\begin{figure}[!ht]
    \centering
    \subfloat[Site 1]{{\includegraphics[width=5.15cm]{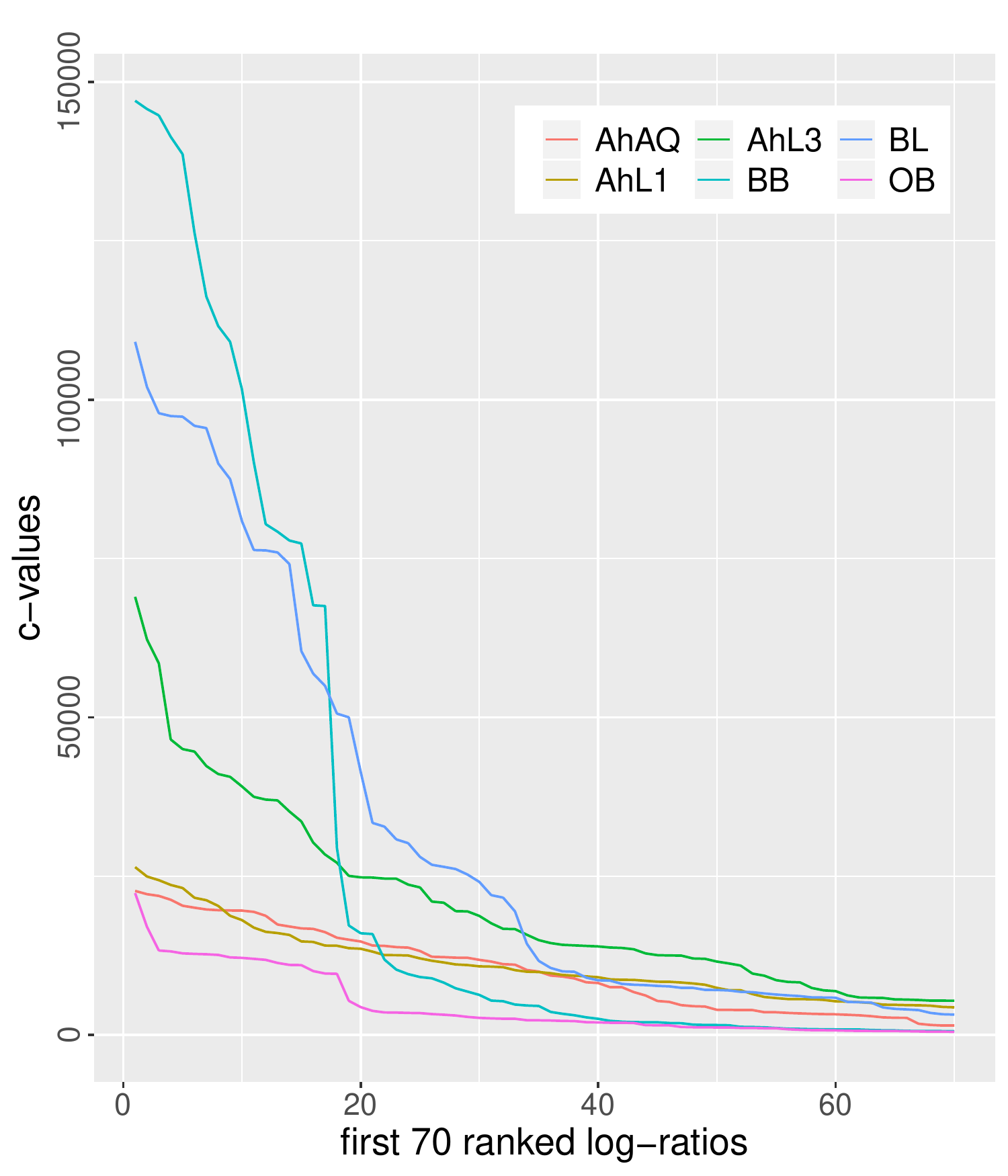} }}%
    \subfloat[Site 2]{{\includegraphics[width=5.15cm]{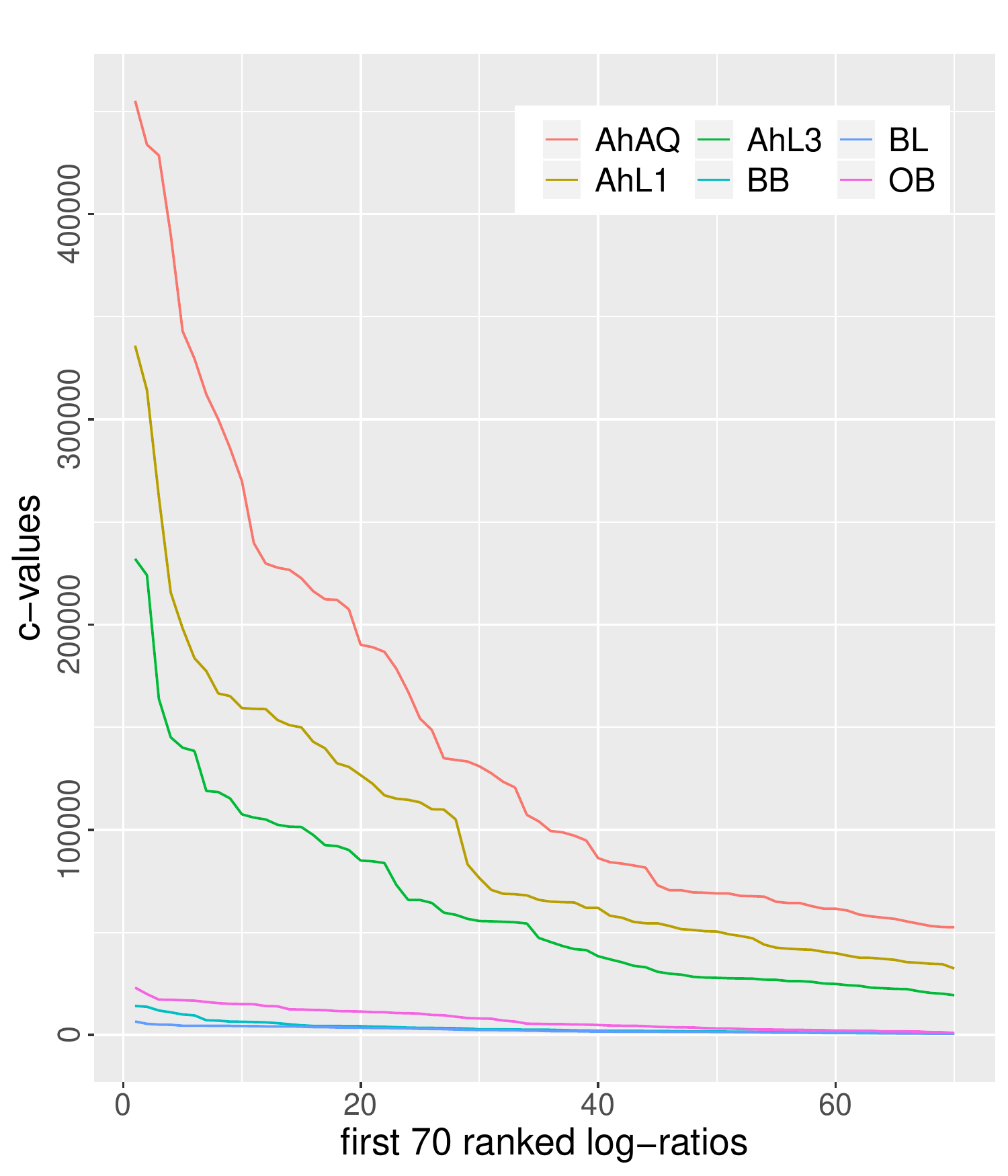} }}
    \subfloat[Site 3]{{\includegraphics[width=5.15cm]{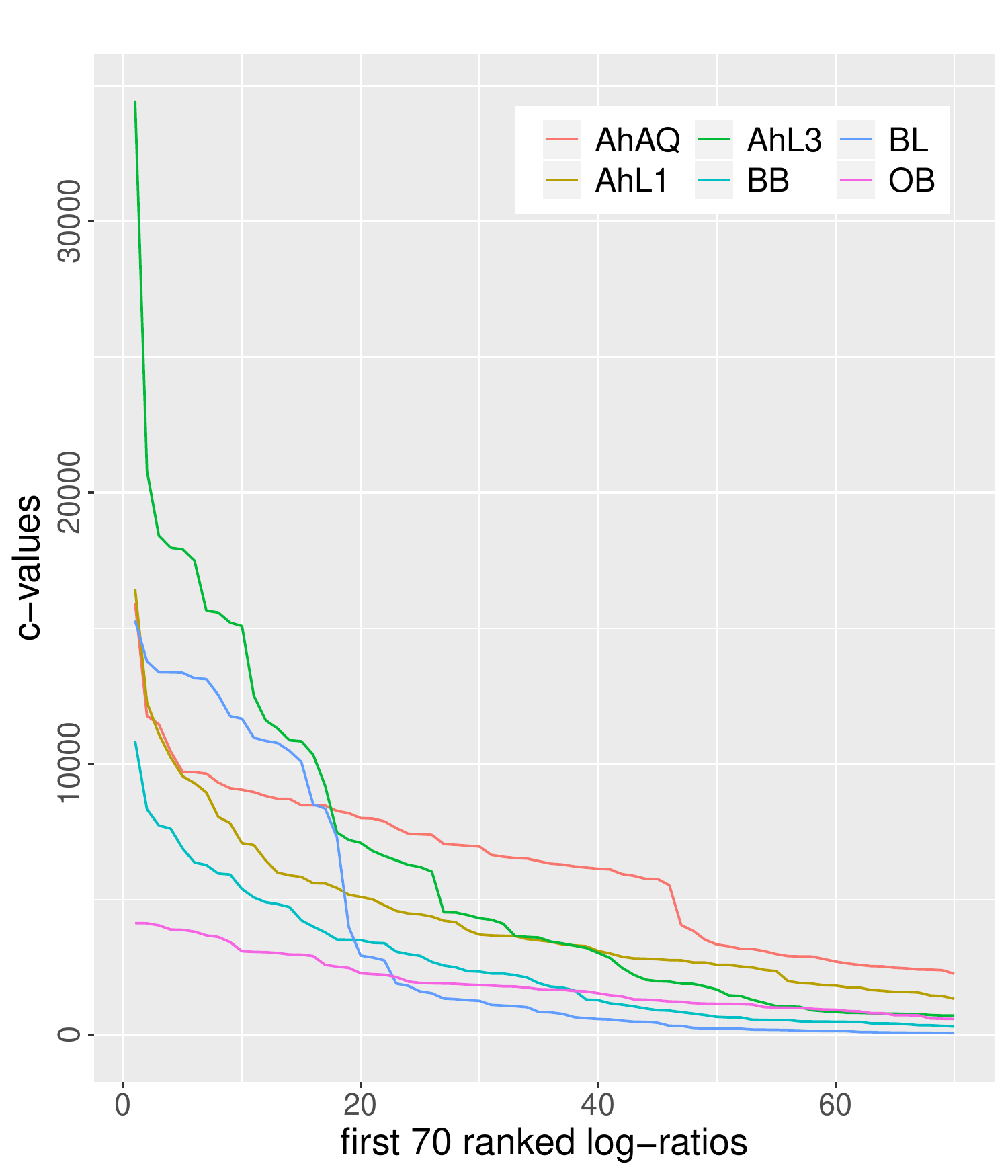} }}%
    \caption{Top-ranked 70 $c$-values, computed for 6 different sample materials
    and the three different sample sites.}%
    \label{fig:lines_brgm}%
\end{figure}

Figure~\ref{fig:lines_brgm} presents for each sample site the top-ranked 70
$c$-values from all pairwise log-ratios of the GAM fits, separated by 
sample material. Since the log-ratios of the fitted values are scaled 
to the interval $[0,1]$, the $c$-values are comparable, 
regardless of sample site and sample material. We obtain the highest $c$-values 
for Site 2, which is the most reliable sample site
due to the higher number of observations. The plot for Site 2 reveals a clear difference in the top ranked $c$-values for the 
mineralization, while the soils seem to be highly informative.
All sites show that sample material OB performs worst in terms of the $c$-values,
and thus this is the least interesting sample material.

\begin{figure}[!ht]
    \centering
    \subfloat[Ah horizon with Aqua Regia leach]{{\includegraphics[page = {3}, width=7.1cm]{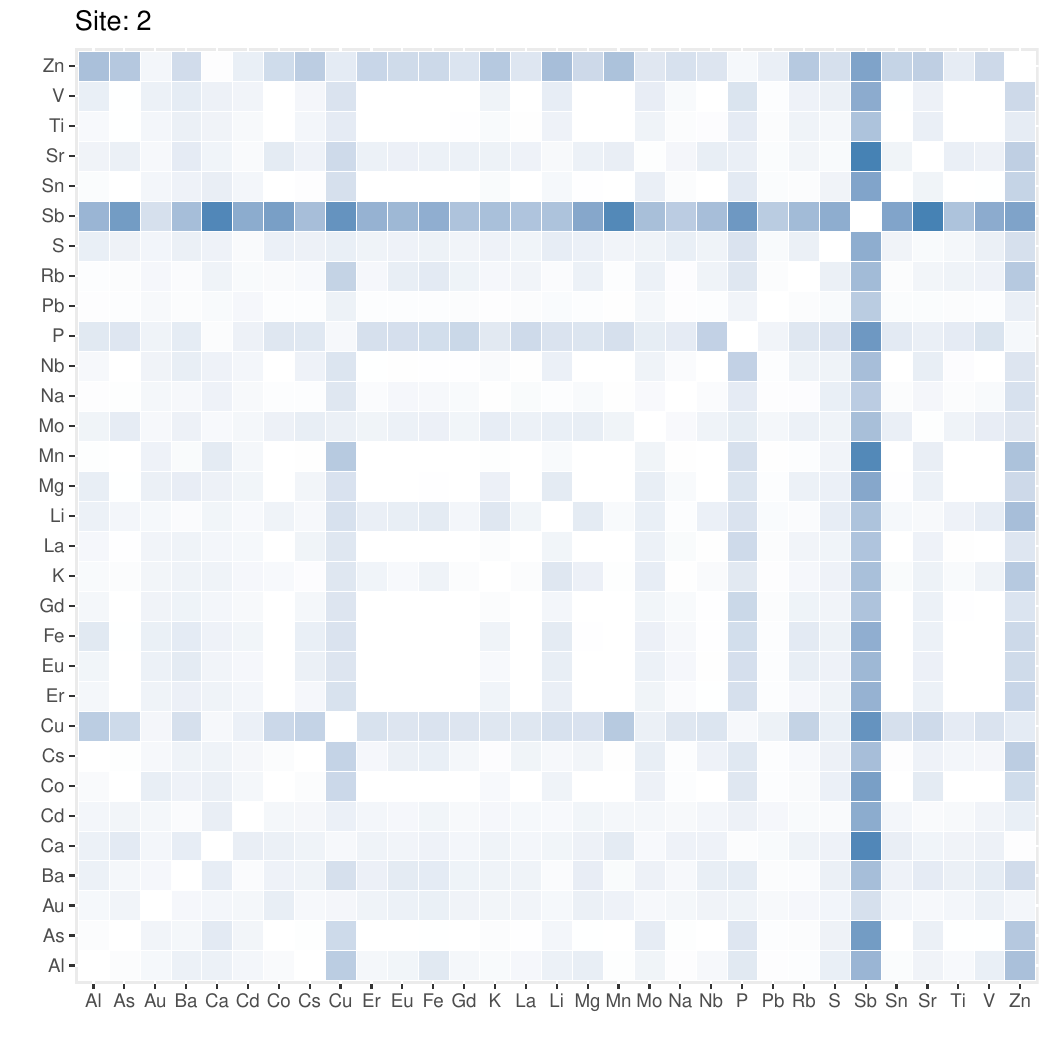} }}%
    \qquad
    \subfloat[Bramble branch]{{\includegraphics[page = {3}, width=7.1cm]{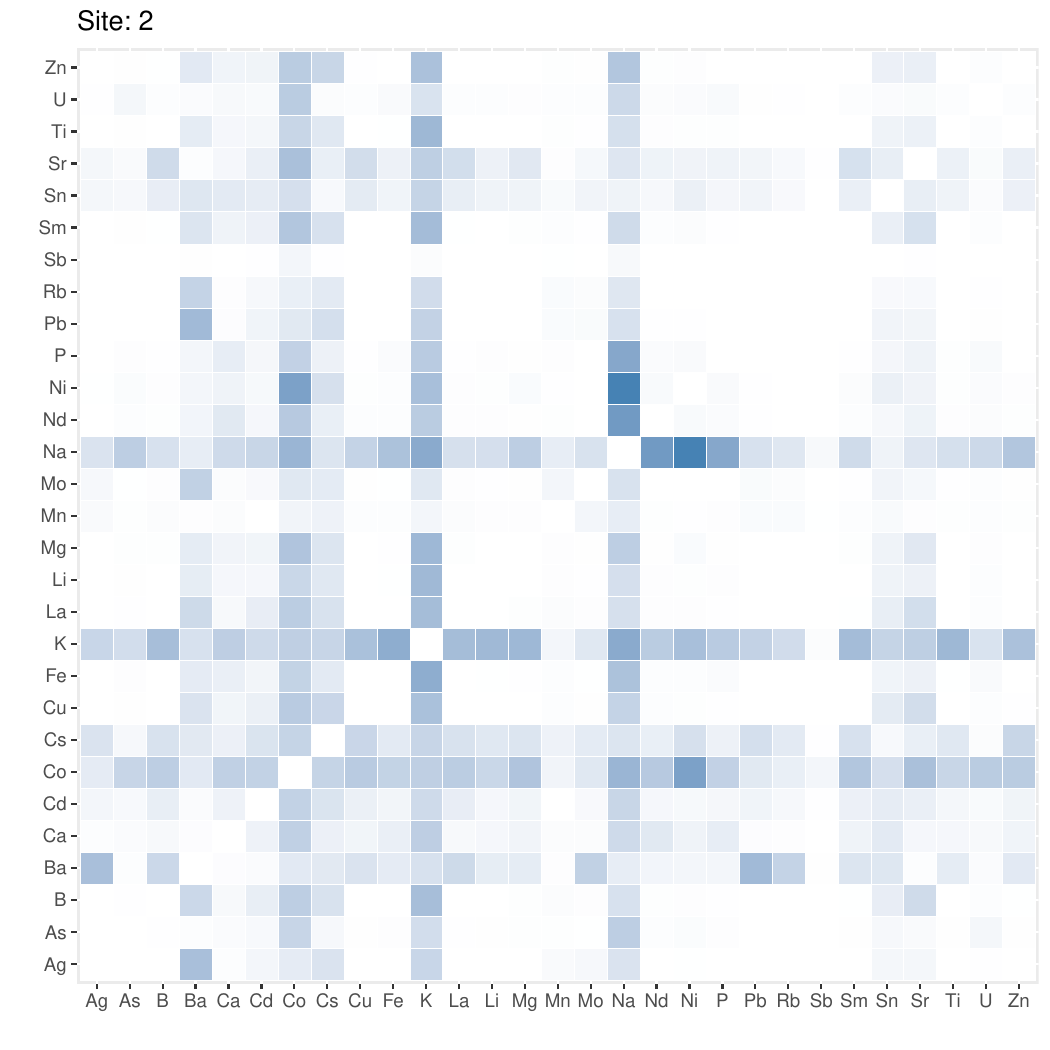} }}%
    \caption{Heatmaps of the $c$-values for the BRGM data - soil and plant material.}
    \label{fig:heat_brgm}%
\end{figure}

The heatmaps in Figure~\ref{fig:heat_brgm} confirm our findings.
The left plot for the soil material AhAQ identifies Sb (and to a lesser
extent Zn) as important pathfinder element of mineralization.
The right plot for plant BB uses the same color scheme, but 
represents much lower $c$-values (see Figure~\ref{fig:lines_brgm}, middle).
This heatmap shows a rather inhomogeneous structure and thus no 
clear pathfinder elements.

\section{Summary}

Due to the technological developments, mineral exploration nowadays belongs
to the most important tasks in geochemistry. Although many chemical 
elements can be investigated for their concentration in different sample
materials, sampling is still time- and cost-intensive, and this is the
reason why usually only 20-60 samples are available at a potentially
mineralized zone. The common strategy is to position the samples on (a)
linear transect(s), crossing the mineralized zones, and mineralization would 
then appear in terms of increased element concentrations.

Rather than investigating single element concentrations, \citet{miksova}
have developed a method based on considering log-ratios of all pairs
of elements. Since the number of possible pairs increases quickly with
the number of investigated chemical elements, a strategy has been proposed
to rank the element pairs according to their relevance for mineral exploration.
This strategy uses a measure of curvature for log-ratios of smooth fits of the concentration
values. The resulting $c$-values are normalized and can be compared across
different pairs, and even across different sample materials and sites.

In this paper we have demonstrated the usefulness of this procedure based
on two data sets that have been collected specifically for the purpose of
mineral exploration. For the first data set originating from Greenland it
has been shown that the $c$-values indeed identify important pathfinder
elements to confirm presumed mineralized zones, but they seem also 
promising to point at new locations with potential mineralization.
The second data set from France was employed to investigate which
sample material is most promising to detect mineralization.
It turned out that the soil samples are much more informative than the
plant samples, but this may again depend on the type of mineralization,
and probably even on further factors.

In our future work we will extend the methodology of \citet{miksova} to
the case where the samples are not necessarily taken along a linear
transect, but on a sample grid with different $x$- and $y$-coordinates. 
This means
that the smooth fits as well as the curvature measure need to be extended to
the two-dimensional case.

\section*{Acknowledgments}

The authors acknowledge support from the UpDeep project
(Upscaling deep buried geochemical exploration techniques into European
business, 2017-2020), funded by the European Information and Technology Raw Materials. We thank GTK personnel Janne Kivilompolo and Kari Kivilompolo for mountain birch sampling and Jukka Konnunaho for organizing the funding and project management of Mineral potential of northern Finland.

\vspace{1cm}

\bibliographystyle{plainnat}
\bibliography{ref}

\end{document}